\documentclass[aip, jap, reprint]{revtex4-1}

\usepackage [english]{babel}
\usepackage{graphicx}
\graphicspath{{pics//}}

\usepackage{natbib}

\begin{document} 

\title{The dynamics of spontaneous hydrogen segregation in LaFe$_{13-x}$Si$_x$H$_y$}

\date{\today}

\author{Oliver L. Baumfeld}
\email[Electronic mail: ]{o.baumfeld12@imperial.ac.uk}
\affiliation{Department of Physics, Blackett Laboratory, Imperial College London, London SW7 2AZ, United Kingdom}

\author{Zsolt Gercsi}
\affiliation{Department of Physics, Blackett Laboratory, Imperial College London, London SW7 2AZ, United Kingdom}
\affiliation{CRANN and School of Physics, Trinity College Dublin, Dublin 2, Ireland}

\author{Maria Krautz}
\affiliation{Leibniz Institute for Solid State and Materials Research Dresden,  01069 Dresden, Germany}
\affiliation{Institute for Materials Science, Dresden University of Technology, 01069 Dresden, Germany}

\author{Oliver Gutfleisch}
\affiliation{Materials Science, Technical University Darmstadt, 64287
Darmstadt, Germany}

\author{Karl G. Sandeman}
\affiliation{Department of Physics, Blackett Laboratory, Imperial College London, London SW7 2AZ, United Kingdom}

\begin{abstract}
By means of time- and temperature-dependent magnetization measurements, we demonstrate that the timescale of hydrogen diffusion in partially-hydrogenated LaFe$_{13-x}$Si$_x$H$_y$ is of the order of hours, when the material is held at temperatures close to its as-prepared Curie temperature, $T_{C0}$. The diffusion constant is estimated to be $D \approx 10^{-15}$ - $10^{-16}$\,m$^2$s$^{-1}$ at room temperature. We examine the evolution of a magnetically phase-separated state upon annealing for 3 days at a range of temperatures around $T_{C0}$, and show that the thermodynamic driving force behind hydrogen diffusion and phase segregation may be attributed to the lower free energy of hydrogen interstitials in the ferromagnetic state relative to the paramagnetic state.
\end{abstract}

\maketitle

\section{Introduction}
Magnetocaloric effect (MCE) research at room temperature is driven by the future application of magnetic refrigeration as  a cleaner and more energy efficient alternative to common gas compression-based cooling. The MCE is characterized in terms of either an isothermal entropy change $\Delta S$ or an adiabatic temperature change $\Delta T_{ad}$ and magnetic refrigerant studies typically focus on how to improve these two figures of merit.

One of the principal candidate magnetic refrigerants for room temperature magnetic cooling technology is based on LaFe$_{13-x}$Si$_x$ (LaFeSi).~\cite{Sandeman2012} This material crystallizes in the cubic, NaZn$_{13}$-type structure and undergoes a paramagnetic (PM) to ferromagnetic (FM) phase transition at a temperature, $T_C$, that depends on the silicon content.~\cite{Palstra1983}  Below a silicon content of $x \approx 1.8$, the phase transition is first order with a large MCE. An itinerant electron metamagnetic transition can be induced by a magnetic field above $T_C$ but below a metamagnetic transition temperature $T_M$.~\cite{Fujita1999, Hu2001, Fujita2001, Fujieda2002} By increasing the silicon content $T_C$ increases while $T_M$ decreases and the transition becomes less first order until the metamagnetic transition line merges with the magnetic transition line at $x \approx 1.8$.\cite{Fujita2001} For $x > 1.8$ the magnetic transition is second order with a much smaller MCE and no metamagnetic transition can be found. The transition temperature ranges from 195\,K for $x = 1.55$ to 245\,K for $x=2.2$.

In order to utilize the giant MCE in LaFeSi for magnetic cooling at room temperature, the low transition temperature needs to be overcome.  Although increasing the silicon content increases $T_C$, the FM-PM transition is broadened and the MCE is drastically reduced.\cite{Fujieda2002, Fujita2003}  Other means of tuning $T_C$, including the replacement of La or Fe by various elements such as Ce, Co, Ni, Mn, have been tested, along with the introduction of interstitial atoms such as B, C, N and H.\cite{Hu2002, Wang2003, Liu2003, Chen2003, Teixeira2012}  The most promising way of tuning $T_C$ to room temperature is currently the insertion of interstitial hydrogen because the favorable characteristics of the magnetic transition are preserved.\cite{Fujieda2001, Chen2003a, Fujita2003, Jia2008a} The presence of hydrogen only increases the lattice parameters, thereby changing the exchange interaction but it changes the density of states of the 3$d$-bands of iron by only a small amount.\cite{} The entropy change $\Delta S$ and the adiabatic temperature change $\Delta T_{ad}$ remain high while the thermomagnetic hysteresis is still small, due to the insensitivity of the electronic structure to the hydrogenation.\cite{Jia2008a} Hydrogenation increases $T_C$ from about 200\,K to 330\,K as the hydrogen content $y$ is varied from 0 to 1.75.\cite{Fujita2003}

For application purposes, complete control of $T_C$ and the MCE is desired.  However, it has recently been discovered that the single, sharp magnetic phase transition of partially hydrogenated LaFe$_{13-x}$Si$_x$H$_y$ ($y < 1.75$) is unstable.\cite{Barcza2011, Krautz2012, Zimm2013} The transition has been seen to split into two separate transitions with distinctive $T_C$s when the material is kept close to its unsplit phase transition temperature, $T_{C0}$. Such a split of $T_C$ implies that a diffusion of hydrogen atoms has taken place within the material, leading to hydrogen segregation and hence an inhomogeneous distribution of $T_C$s.

Thus far, the hydrogen diffusion has been thought to be driven by the coexistence of PM and FM magnetic phases in the vicinity of a first order $T_C$.  In the simplest model, diffusion of the hydrogen atoms would take place from the lower volume PM phase to the higher volume FM phase~\cite{Barcza2011, Krautz2012} since the ferromagnetic unit cell volume is about 1 to 2\% larger while the symmetry of the cubic crystal structure is preserved. \cite{Fukamichi2000, Fujita2001, Jia2006c}  As a result, the $T_C$ of the PM regions $T_{C1}$ decreases while the $T_C$ of the FM regions $T_{C2}$ increases, stabilizing the resulting states and eventually leading to the appearance of two distinctive magnetic phase transitions.

When the material is held at temperatures well above or below the initial $T_{C0}$ the hydrogen segregation can be reversed and the initial, homogeneous state can be recovered. It has been reported that the time for this thermally activated recombination ranges from days to minutes as the recombination temperature is increased.\cite{Krautz2012, Zimm2013}

While the dynamics of recombination have been investigated, little has been done to investigate the timescale of the transition splitting process which is of great relevance to the efficiency of cooling cycles (as any change in the apparent $T_C$ of a refrigerant would have a negative impact on the performance of a device.\cite{Zimm2013}) Splitting has been reported for samples with a completed diffusion process i.e.\ after keeping the sample near $T_{C0}$ for 1 or 2 months. In this work we further investigate the dynamics of the hydrogen diffusion in LaFe$_{13-x}$Si$_x$H$_y$ at intermediate time scales, focusing on the splitting process.

\section{Experimental Details}
All of our experiments have been conducted on a single powder sample of LaFe$_{11.6}$Si$_{1.4}$H$_{1.3}$ with a particle size smaller than 100\,$\mu$m. Details of the sample preparation are given elsewhere. \cite{Krautz2012}
Magnetization measurements were performed using a commercial vibrating sample magnetometer (Quantum Design PPMS). Since the magnetic state of the sample strongly depends on its annealing history, the latter will be given in detail. Between annealing protocols, the sample was heated well above $T_C$, typically to 390\,K at 1\,K/min and subsequently cooled down~\footnote{Comparing measurements performed 12 months apart we observed a decrease of $T_{C0}$ in the unsplit, recombined state of about 2\,K indicating a slight loss of about 0.03 hydrogen atoms per unit cell. The loss of hydrogen might have been thermally activated during the heating to 390\,K between the measurements assisted by the low pressure of about 10\,mbar in the sample space.} to reverse any previous diffusion and to ensure that the hydrogen is uniformly distributed as at this temperature the recombination speed is in the range of minutes.\cite{Krautz2012} We hereafter call this the recombined state. The sample was cooled down to the temperature of interest, $T_A$, and kept at $T_A$ for a period of time lasting between several hours and 3 days. Subsequently the magnetization was measured from 270\,K to 320\,K with a sweep rate of 1\,K/min in a low magnetic field of 100\,Oe.

\section{Results and Discussion}

\begin{figure}
	\includegraphics[width=0.95\columnwidth]{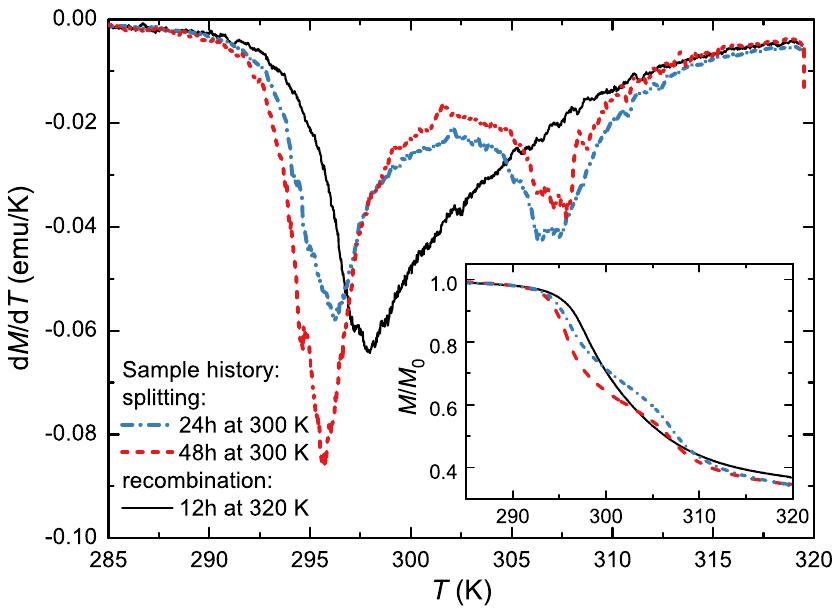}
	\caption{Temperature dependent normalized magnetization (inset) and its temperature derivative of LaFe$_{11.6}$Si$_{1.4}$H$_{1.3}$ taken in a magnetic field of 100\,Oe on heating. The blue and red curves show a split state where, starting from the recombined state, the sample was kept at $T_A = 300$\,K for 24\,h and 48\,h, respectively. The black curve shows the partially recombined state where, after starting from a split state, the sample was kept at 320\,K for 12h.}
	\label{fig:MvsT}
\end{figure}

Figure~\ref{fig:MvsT} shows the temperature dependent magnetization and its temperature derivative for LaFe$_{11.6}$Si$_{1.4}$H$_{1.3}$ after two splitting protocols ($T_A = 300$\,K for 24\,h and 48\,h) and one recombination protocol (320\,K for 12\,h). The transition temperature $T_{C0}$ of the sample in the recombined state is around 296\,K on heating, where $T_C$ is defined as the peak maximum of the temperature derivative. As in previous measurements of this material, \cite{Barcza2011, Krautz2012, Zimm2013} the single sharp transition decomposes on splitting into two separate transitions with two distinctive $T_C$s which corresponds to a splitting of the peak in the derivative of the magnetization. While in previous reports the ``peak splitting'' was observed after months, here we show that a significant splitting occurs after only 24 hours.

\begin{figure}
	\includegraphics[width=1\columnwidth]{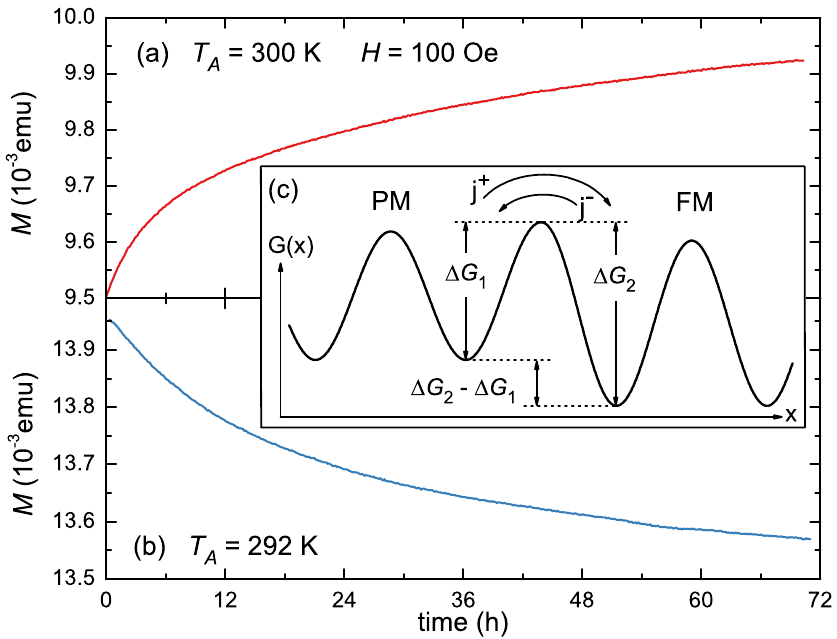}
	\caption{Time dependent small field magnetization of LaFe$_{11.6}$Si$_{1.4}$H$_{1.3}$ at (a) $T_A = 292$\,K  and (b) $T_A = 300$\,K. At $t=0$ the sample has been in the recombined state. The inset (c) shows a schematic free energy landscape at a PM-FM boundary where each minimum represents a hydrogen site. Hydrogen migration across the magnetic phase boundary occurs with the hopping rates $j^+$ and $j^-$ where $\Delta G_{1}$ and $\Delta G_{2}$ are the corresponding activation energies.}
	\label{fig:MVSTIME}
\end{figure}

In Fig.~\ref{fig:MVSTIME} we show the time dependent magnetization for $T_A$ = 292\,K and 300\,K. 
The magnetization changes can be explained by the decomposition of $T_{C0}$ in $T_{C1}$ and $T_{C2}$ due to hydrogen diffusion from the PM regions into the FM regions.
$T_{C1}$ of the PM regions shifts to lower temperatures as their hydrogen content decreases which causes their magnetization to decrease. Likewise the magnetization of the FM regions increases as $T_{C2}$ increases as they are now closer to saturation. 
The overall magnetization, as shown in Fig.~\ref{fig:MVSTIME}, is then given by the sum of the magnetization of the PM and FM regions.
For $T_A = 292$\,K, which is below $T_{C0}$, the PM regions are predominant and the magnetization decreases. For $T_A = 300$\,K, i.e.\ above $T_{C0}$, the FM regions are predominant and the magnetization increases.
The rate of the time dependent magnetization change serves as an indicator for the diffusion process although it is not a direct measurement of the diffusion rate. The magnetization in Fig.~\ref{fig:MVSTIME} (a) and (b) does not follow an exponential and a physically meaningful fit was not found. To estimate the diffusion constant at room temperature we define a characteristic timescale $\tau$ as the time at which the magnetization is midway between its value at 0 and 72 hours which gives $\tau \approx 15$\,hours.
We also note that the slope of the magnetization has significantly decreased after 3 days but it has not saturated. Therefore it can be assumed that there is still some ongoing diffusion. The similarity of the time constants of the 292\,K and 300\,K magnetization curves is remarkable but its origin is unclear given the complicated temperature and history dependence of the magnetization.

With a tentative characteristic timescale of $\tau \approx 15$\,h and a characteristic magnetic domain size of $l \approx$~25-30\,$\mu$m, derived from imaging experiments,~\cite{Krautz2012} the diffusion constant $D$ can be estimated by considering the diffusion equation in the two limiting cases of homogenization in an infinite periodic sample and diffusion in a semi-infinite sample.\cite{Porter1992}
In the first case it is useful to Fourier expand the concentration profile with the $n$th Fourier component of each dimension having a time dependent amplitude of $A = \exp(-tD\pi^2n^2/l^2)$, where $2l$ is the wavelength of the first harmonic. For a periodic concentration profile higher harmonics can be neglected and for 3 dimensions the point of half amplitude is given by $A = 0.5 = (\exp(-tD\pi^2/l^2))^3$. Rearranging gives $D = 0.0234 \times l^2/t$ which leads to $D \approx 2.5\times10^{-16}$\,m$^2$s$^{-1}$.
For a semi-infinite sample the concentration profile is proportional to the error function erf$(x/2\sqrt{Dt})$. Since erf(0.5)~$\approx$~0.5, the depth at which the hydrogen concentration is midway is given by $(x/2\sqrt{Dt}) \approx 0.5$. If we assume that for $\tau=15$\,h the depth is half the domain size $l$ then $D = 0.25 \times l^2/t$ which leads to $D \approx 2.9\times10^{-15}$\,m$^2$s$^{-1}$.
For particles smaller than 100\,$\mu$m the geometry of the real sample is somewhere between those two limits and so must be the diffusion constant which is then approximately $D \approx 10^{-15}$ - $10^{-16}$\,m$^2$s$^{-1}$ near room temperature.
Due to the dimension of the diffusion constant D (length$^2$/time) the error or variation in the characteristic magnetic domain size dominates in comparison to the error in the characteristic timescale $\tau$.
As shown by Zimm et al.~\cite{Zimm2013} the speed of the reverse process is temperature dependent and might be described by an Arrhenius-type equation. It is reasonable to assume a similar temperature dependency for the peak splitting process. Since the speed of diffusion is dependent on the number of unoccupied sites it can be expected that diffusion is slowest for materials close to full hydrogenation.

\begin{figure}
	\includegraphics[width=0.95\columnwidth]{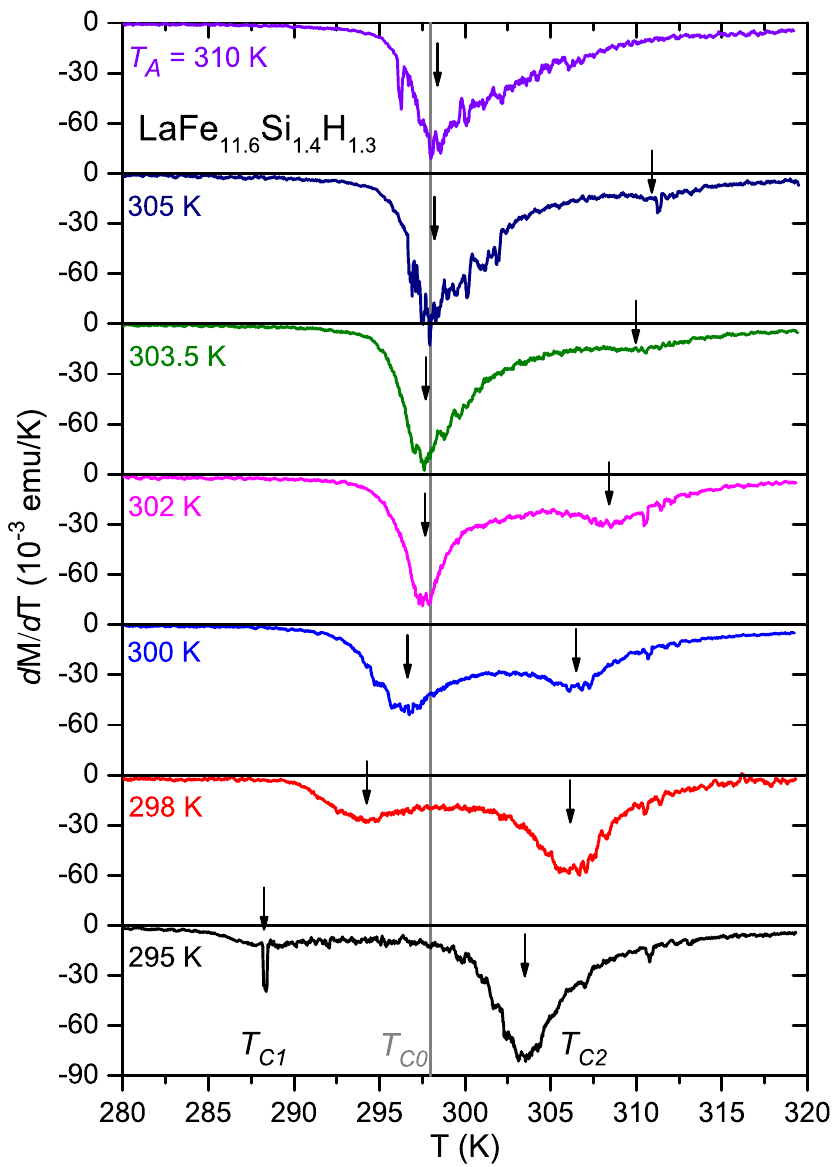}
	\caption{Temperature derivatives of the normalized magnetization after the sample has previously been in the recombined state and was afterwards kept for 3 days at $T_A = 310$, 305, 303.5, 302, 298, 295 and 292\,K and after 2 days at $T_A = 300$\,K showing the development of the two phases. The left and right black arrows indicate $T_{C1}$ and $T_{C2}$, respectively and the gray line at 296\,K indicates $T_{C0}$.}
	\label{fig:dMdT}
\end{figure}

Temperature dependent magnetization measurements were taken after keeping the recombined sample for 3 days at several values of $T_A$ around $T_{C0}$. The results are presented in Fig.~\ref{fig:dMdT} as temperature derivatives of the magnetization.
In the 310\,K curve no second peak can be distinguished and the magnetic transition looks similar to that found in the recombined state shown in Fig.~\ref{fig:MvsT}.
By going to lower annealing temperatures a small and broad second peak emerges with a median above 310\,K. This peak, corresponding to $T_{C2}$ of the FM regions present at $T_A$, is shifted to lower temperatures while increasing in size. The increase in size is attributed to the increasing percentage of FM regions as $T_A$ approaches $T_{C0}$. At lower $T_A$ the FM phase becomes the majority phase and $T_{C2}$ shifts towards $T_{C0}$. A similar behavior can be found for the lower temperature peak corresponding to $T_{C1}$ of the PM regions present at $T_A$ as $T_A$ is increased.
Because of our definition of $T_C$ as the maximum of $dM/dT$ and because of the asymmetry of the transition, the annealing temperature at which the peaks of $T_{C1}$ and $T_{C2}$ are of the same size is not $T_{C0}$. The relevant $T_A$ is closer to 300\,K and corresponds instead to where the magnetization is halfway between its FM and PM value.

The separation between the peaks is limited by the thermodynamic driving force of the diffusion. If this force is attributed to a gradient in the hydrogen concentration $n$ per unit cell volume $V$ then, in equilibrium, it must be the same in the PM and FM phase and hence $n_{PM}/n_{FM} = V_{PM}/V_{FM}$. Typical equilibrium values of the hydrogen concentration in the sample can be estimated from $T_{C1}$ and $T_{C2}$ to be $n_{PM}\sim1.2$ and $n_{FM} \sim 1.4$ which gives  $n_{PM}/n_{FM} = 0.86$. This would correspond to a 14\% decrease of the volume in the PM phase compared to the FM phase whereas the experimentally determined decrease is about 1 to 2\%.\cite{Fujita2001, Jia2006c, Fujieda2001}
In the following we will show that an atomic jump process model~\cite{Wert1950, Vineyard1957} with different activation energies $\Delta G_{1}$ and $\Delta G_{2}$, as shown in Fig.~\ref{fig:MVSTIME} (c), would be a reasonable origin of the driving force. We restrict our analysis to classical thermal activation although for the lightweight hydrogen, quantum corrections can not be ruled out even at room temperature.

Neutron diffraction measurements on LaFe$_{13-x}$Si$_{x}$H$_{y}$ compounds have shown that the hydrogen occupies either the 24\,d or 48\,f position inside a 2La--4Fe octahedron.\cite{Fujieda2008, Rosca2010} The spontaneous magnetostriction below $T_C$ increases the La-H and Fe-H distances and it is reasonable to assume that this changes the free energy of the hydrogen atom on its interstitial sites. This means that the energy barrier for a hydrogen atom jumping from a PM to a FM site, $\Delta G_{1}$, is different from the energy barrier for jumping from a FM to a PM site, $\Delta G_{2}$. The jump rate from the PM to the FM site, $j^+$, is given as
\begin{equation}
j^+ = \nu_{1} \cdot n_{PM} \left(n_{max}-n_{FM}\right) \cdot \exp\left( \frac{-\Delta G_{1}}{ k_B T} \right)
\end{equation}
where $\nu_{1}$ is the number of jumps attempted per second, $n_{PM}$ denotes the occupancy of the PM site and $(n_{max}-n_{FM})$ is proportional to the probability that the FM site is unoccupied. The jump rate in the opposite direction $j^-$ is defined accordingly. The principle of detailed balance states that in equilibrium the jump rates across each energy barrier are equilibrated and therefore $j^+=j^-$. This leads to the following relation for the free energy difference $\Delta G = \Delta G_{2} - \Delta G_{1}$:
\begin{equation}
	\Delta G = k_B T \cdot \ln \left( \frac{\nu_{2}}{\nu_{1}} \cdot
	 \frac{ n_{FM} \left(n_{max}-n_{PM}\right)}{n_{PM} \left(n_{max}-n_{FM}\right)}\right).
\end{equation}
At 300\,K and in equilibrium (at large $t$) the hydrogen occupations for this sample are given by $n_{PM} = 1.20(5)$, $n_{FM} = 1.40(5)$ and $n_{max} = 1.75(10)$.\cite{Krautz2012} The jump attempts per second $\nu_{1}$ and $\nu_{2}$ should be proportional to the Debye frequency which depends only slightly on the volume and their ratio is therefore close to one. Using these values the free energy difference is given as $\Delta G = 15(6)$\,meV. This is not incompatible with the tentatively estimated activation energy in the PM phase of 440\,meV.\cite{Zimm2013}
Preliminary density functional theory calculations show that a hydrogenated PM phase is in equilibrium with a FM phase of higher hydrogen concentration, corroborating a free energy argument.~\cite{Gercsi2013}

\section{Conclusion}
Our measurements demonstrate that the timescale of hydrogen diffusion is faster than initially reported. We have found that a significant part of the hydrogen diffusion in LaFe$_{13-x}$Si$_x$H$_y$ happens within hours but that there are still measurable diffusion-related magnetization changes after 3 days. The diffusion constant at room temperature is estimated to be $D \approx 10^{-15}$ - $10^{-16}$\,m$^2$s$^{-1}$.
This is at the slow end of the diffusion of small concentrations of interstitial hydrogen in metals which typically ranges from $10^{-8}$ to $10^{-15}$\,m$^2$/s at room temperature but it is faster than the diffusion of heavy interstitials such as C and N.\cite{Volkl1975}
We have examined the range of annealing temperatures around $T_{C0}$ where a significant splitting of the transition can be expected and the evolution of the magnetization within this temperature window. 
Nonetheless a large thermodynamic driving force is present in the two-phase region around $T_{C0}$. The origin of this force may be explained by a free energy difference of the hydrogen atoms in the PM and FM unit cell rather than solely by the PM-FM volume change.

\begin{acknowledgments}
The research leading to these results has received funding from the European Community's 7th Framework Programme under Grant agreement 310748 ``DRREAM''.  Financial support is acknowledged from The Royal Society (KGS), EPSRC grant EP/G060940/1 (KGS and ZG), and an EPSRC DTG studentship is acknowledged by OLB. We are grateful to L.\ F.\ Cohen for useful discussions.
\end{acknowledgments}

%

\end{document}